\documentclass[
 aip,
 amsmath,amssymb,
 reprint,%
]{revtex4-1}

\usepackage{graphicx}% Include figure files
\usepackage{dcolumn}% Align table columns on decimal point
\usepackage{bm}% bold math
\usepackage{caption}
\usepackage{subcaption}
\usepackage[utf8]{inputenc}
\usepackage[T1]{fontenc}
\usepackage{mathptmx}
\usepackage{hyperref}

\usepackage{color,soul}
\newcommand{\cblack}{\color{black} }

\begin{document}

\preprint{AIP/123-QED}

% Force line breaks with \\
\title{Statistical mechanics study of the introduction of a vaccine against COVID-19 disease.}
\author{Hilla\ De-Leon}
 \email[E-mail:~]{hdeleon@ectstar.eu}
 \affiliation{INFN-TIFPA Trento Institute of Fundamental Physics and Applications, Via Sommarive, 14, 38123 Povo TN, Italy}
 \affiliation{
 European Centre for Theoretical Studies in Nuclear Physics and Related Areas (ECT*),
 Strada delle Tabarelle 286, I-38123 Villazzano (TN), Italy}
 \author{Francesco Pederiva}
\email[E-mail:~]{francesco.pederiva@unitn.it}
 \affiliation{INFN-TIFPA Trento Institute of Fundamental Physics and Applications, Via Sommarive, 14, 38123 Povo TN, Italy}
 \affiliation{Dipartimento di Fisica, University of Trento, via Sommarive 14, I–38123, Povo, Trento, Italy}

\date{\today}% It is always \today, today,
 % but any date may be explicitly specified

\begin{abstract}

By the end of 2020, a year since the first cases of infection by the Covid-19 virus have been reported, there is a light at the end of the tunnel.  Several pharmaceutical companies made significant progress in developing effective vaccines against the Covid-19 virus that has claimed the lives of more than a million people over the world. On the other hand, there is growing evidence of re-infection by the virus, which can cause further outbreaks.
In this paper, we apply statistical physics tools to examine the vaccination rate required to control the pandemic for three different vaccine efficiency scenarios.  Also, we study the effect of temporal restrictions/reliefs on the pandemic's outbreak, assuming that re-infection is possible.
When examining the efficiency of the vaccination rate of the general population in preventing an additional outbreak of the disease, we find that a high vaccination rate (where at least 0.3\% of the population is vaccinated daily, which is equivalent to $\approx$ 1 million vaccine doses in the United States daily) is required to gain control over the spread of the virus without further restrictions. Due to feasible limitations on the vaccination rate, the vaccination process should be accompanied by the prevailing restrictions until most of the population is vaccinated.
\end{abstract}

\maketitle

\section{Introduction}
 
 %\section{}\label{MC}

%\cred

% WE NEED THAT?

% Statistical mechanics provides a set of powerful tools to model various biological and medical problems (see, for example, Refs.~\cite{blossey2019computational,buldyrev1998analysis,stanley1994statistical} and many more). One of the most studied these days is pandemics' diffusion, prompted by the current COVID-19 emergency (see, for example, Ref.~\cite{SOHRABI202071}). Also, many current researchers study the physical aspects of the virus's spreading (see, for example, Refs. \cite{doi:10.1063/5.0012009,doi:10.1063/5.0011960,doi:10.1056/NEJMc2004973,doi:10.1063/5.0015984,doi:10.1063/5.0015984}). 
% Many techniques currently employed are based on the solution of differential equations (see, for example, Refs.~\cite{karin2020adaptive,prem2020effect,zhao2020modeling} and many more) or fitting formulae (see, for example, Refs.~\cite{bliznashki2020bayesian,olsson2020ongoing}). Both techniques are based upon varied parameters to obtain several scenarios that are then treated as parts of a statistical ensemble to analyze. For example, in many countries (e.g., Germany, Italy), there is an ample discussion about the role of the so-called $R_0$ parameter, i.e., the average number of individuals that a single actively infectious person can pass the virus to. The procedures to estimate $R_0$ are all based on a-posteriori analyses, but are usually part of the parameters that governments use to decide on measures to be taken.
% \cblack
 The COVID-19 disease has been spreading worldwide in the last year, whereby at the end of 2020, tens of millions have been infected from the virus and over a million people have died. 

Today, many companies and research institutes are in an advanced research stage to issue an efficient vaccine that will help to suppress the pandemic, eventually allowing for a faster return to the pre-corona life. As of this writing, at least two companies have reported proven efficacy of the vaccine \cite{pfizer_2020, modernas_2020}. 
As a result, in many countries worldwide, the question now arises about how the vaccine will affect the spread of the epidemic and what an effective rate of population vaccination  will ensure control of the epidemic.
 Moreover, during the time that has passed since the first outbreaks of the epidemic, some evidence of re-infection with the virus has accumulated (see for example Refs.~\cite{GOUSSEFF2020816,duggan2020case,alizargar2020risk,edridge2020seasonal,landi2020predictive,science_2020}, and many more).
 
 The time extent of the coverage provided by a vaccine and the probability of re-infection after being subject to symptoms-free contagion or to the manifest disease have to be treated as separate issues from the medical perspective. However, it is clear that,
 whatever mechanism that changes a person's status from potentially covered from the disease to vulnerable, such time extent is a crucial parameter to establish whether an equilibrium condition with a controlled or zero level of infected people can be reached, avoiding another sequence of global outbreaks. 
 
 Closely related to this mechanism, there are also two other parameters: the first is the actual effectiveness of the vaccine
 (claimed at present to be of order 90\%), and the second is the actual availability, which can be translated into the number of people who can be vaccinated in a given amount of time.

 In this paper, we address the study of the influence of such parameters on the evolution of the pandemics using basic principles of statistical mechanics similar to the method presented in Ref.~\cite{de2020particle}, where the population is modeled as a set of interacting classical particles. Each 'particle' can be in four states relative to the health status: susceptible of infection, infected, recovered/died (with 1\% rate of death), and vaccinated. For the recovered people, we assume that re-infection can occur in periods ranging from 150 days from initial infection to one year later. Also, we assume that the vaccine efficiency varies from half a year (180 days) to two years, which is equivalent to infinity in the model's terms. All simulations were tested for different population densities, allowing us to apply the model to very different situations, ranging from a city suburb population to a single university classroom. 
 
 The algorithm is based on standard Monte-Carlo (MC) procedures of sampling the transition among the following states, which are sampled from a statistical distribution, in the spirit of transport MC algorithms.

 Similarly to what reported in Ref.~\cite{de2020particle}, we are using a "one-way" Ising-model Monte-Carlo such that a healthy person (i) can become sick with a daily probability,
 $P_i=\sum_jP_{ij}$, where $P_{ij}$ is a function of the distance between each infected person $(j)$ in the area and the healthy person $(i)$. Based on the epidemiology data, we assume that a sick person stops being sick (i.e., recovered or died) within an average time of $\sim 14$ up to 40 days for the most severe cases, adding this time a finite probability of re-infection and a "vaccination" mechanism which switches a certain fraction of particles from the vulnerable stat to the immune state.
 
 In this approach, the parameter $R_t$, the effective reproduction rate of the virus,\footnote{Note that the onset of the epidemic was characterized by the parameter $R_0$, the basic reproducing number. Now, in many countries, social distance measures are taken that affect the spread of the epidemic. Hence, the calculations in this work were done under the assumption that an effort is made to reduce $R_t$ by various constraints.} is a direct outcome of the simulation and not pre-assumed. This is achieved by the relation between $R_t$ and the doubling time $T_d$, which is a direct result of the infection probability chosen, which is, in turn, a function of observable epidemiological data and the average density on a given area. In this paper, we will present the results of several simulations meant to reproduce the spread of the Coronavirus in the presence of re-infection with and without an effective vaccine. Moreover, the model's high flexibility enables us to control the population density as a function of time and to simulate how temporary changes in $R_t$ affects the virus's spread.
 In Section II, we will describe some details of the model. Section III is devoted to the presentation and discussion of the results, and Sec. IV to the Conclusions.
 
 \section{The parameters and preliminary assumptions}

Given the scarce information available, we had to make some assumptions based on current data, which may be more stringent than it might be required by the real nature of the virus. In this work, similar to Ref.~\cite{de2020particle}, we based on Ref.~\cite{10.7554/eLife.57309} for the Coronavirus epidemiology data and on Refs.~\cite{doi:10.1063/5.0012009,doi:10.1063/5.0011960} for the physical properties of the virus\footnote{We are aware that this model cannot take into account every single spreading event, but since such events affect the initial $T_d$, (which is a function of the population density), and since the infection process is random, we expect that the existence of such events will be reflected in the numerical results. Also, in contrast to real-life, here, there is no time gap between the infection and being tested positive for Coronavirus. Therefore, an immediate decrease in the rate of infection resulting from lockdown is expected in the model in contrast to the real data. \cite{endcoronavirus}.}
 
 All simulations are performed, assuming an initial surface area unit of 1 km$^2$, wherein some simulations we increased/decreased the area as a function of time. The control over the surface area allows to change the simulation population density, which is equivalent to a temporary change of the parameter $R_t$. The use of periodic boundary conditions reduces broad confinement effects (e.g., the lockdown of an entire province or city) and focuses at the \textit{local} dynamics of the infections within that given area \footnote{The application of periodic boundary conditions means that we have an infinite number of identical systems; each system is a replica of the others. I.e., if a person leaves the simulation surface on one side, an equal person will enter the surface from the other side.} \cblack Similar to Ref.~\cite{de2020particle} the population density is a function of the number of households in a certain area since it is crucial to distinguish between the infection among the family and non-household contacts\cite{10.1093/cid/ciaa450}.

 \subsection{Parameterization of the model}
 
 This work has identified a list of parameters and corresponding values that describe the population and the infection's kinetics. 
 \begin{enumerate}
 \item The number of effective households, denoted by $N$.
 \item The fraction of "silent carriers," which have no symptoms (AKA asymptomatic) but can infect other people is denoted by $a_{silent}$, and the probability of transmitting the infection has been set to 0.5.
 \item Each sick person is considered contagious between the 3$^{\text{rd}}$ and the 7$^{\text{th}}$ day.
 \item Simulations are started with a single infected person (the \textit{zero patient}). Also, every day there is a chance of 20\% that a healthy person (who is not recovered/vaccinated) becomes sick without interaction with a known sick person.
 \end{enumerate}

 \subsection{Population dynamics}
 Similar to Ref.~\cite{de2020particle}, our model is based on the principles of Brownian motion, such that for each day, the population position ($R$) and displacement ($\Delta R)$ are given by:
 \begin{equation}
 R\rightarrow R+\Delta R~,
\end{equation}
where $\Delta R=\sqrt{\Delta x^2+\Delta y^2}$ is distributed normally:
\[
 P[\Delta R] =\frac{1}{2\pi^2\sigma^2_{R}} \exp\left(-\frac{\Delta x^2+\Delta y^2}{2\sigma_{R}^2}\right)~,
\]
where $\Delta x$ ($\Delta y$) is displacement in the x-(y-) direction and $\sigma_{R}^2$, the variance, is a function of the diffusion constant, $D$:
\begin{equation}
 \sigma^2_{R}=2Dt~,
\end{equation}
where t=1 day. For a Brownian motion the diffusion coefficient, $D$, would be related to the temperature, $T$, using the Einstein relation:
\begin{equation}
 D=\mu k_{\text{B}}T~,
\end{equation}
where $\mu$ is defined as the mobility, $k_{\text{B}}$ is Boltzmann's constant
and $T$ is the absolute temperature. By fixing $T=1$, the diffusion coefficient would be directly related to mobility. In our model, during all the time period, the population is allowed to move with $\sigma_{R}=500$ meter.
 
 \subsection{The infection probability}
 We assume that this process can be described by a Gaussian function of the distance for each contact with another infected person, weighted with a factor that parametrizes the sick person's conditions and social interaction. \footnote{Since the infection probability is a function of the absolute value of the distance between two people and the standard deviation, several distributions such as a Gaussian distribution and a Lorentzian distribution could serve for modeling the infection probability under the assumption that the simulation's dynamics dependents on $\sigma_r$ and not on the distribution's tail. The choice of Gaussian distribution was since this is the typical distribution for thermal systems approaches for equilibrium, e.g., Maxwell Boltzmann distribution.}

 \begin{eqnarray}\label{eq_pi}
 P_i&=&\text{int}\left(\sum_{j=1}^{n_{sick}}P_{ij} +\xi\right)=\nonumber \\ \\
 &=&\text{int}\left\{\sum_{j=1}^{n_{sick}}\exp\left[\frac{\left(r_i-r_j\right)^2}{2\sigma_r^2}\right]\times f\left(a_{silent},n_{out}\right) +\xi\right\}\nonumber
 \end{eqnarray} 
 
 where:
 \begin{itemize}
 \item $r_i \left(x_i,y_i\right)$ is the location of the $i^{\text{th}}$ healthy person and $r_j \left(x_j,y_j\right)$ is the location of the $j^{\text{th}}$ sick person, so $|r_i-r_j|$ is the distance between them.
 \item $n_{sick}$ is the total number of sick people in the area. 
 \item $\sigma_r$ is the standard deviation (here $\sigma_r=2.4$ meters) %\textcolor{red}{this is very debated, it depends on whether you wear a ask or not, etc. We should add some appropriate citation}
 {since recent studies show that even a slight breeze can drive droplets arising from a human cough over more than 6 meters
~\cite{doi:10.1063/5.0011960}}.
 \item $f(a_{silent},n_{out} )$ is a function that considers the social activity of the sick person and whether he has symptoms, which affects the spread of the virus outside the house. In our model, we estimate that infection by asymptomatic people is approximately 50\% lower than patients with symptoms.
 \item $\xi$ is a number a random number uniformly distributed between 0 and 1, needed for converting to infection probability to a number between 0 and 1.
 
\item We are assuming that each sick person will infect some of his household members. Since the latest estimates are that household infections are $\sim 15\%$ from the known cases (without lockdown, \cite{jing2020household}), we estimated that the number of household infections is are of uniformly distributed between 0 and 3. 
\item During all the runs, we assume that people must wear face-masks so that the daily
infection probability (for the non-household members, eq.~\ref{eq_pi}), is
reduced to \cite{doi:10.1063/5.0015044,19_2020,robinson2020efficacy,eikenberry2020mask,BALACHANDAR2020103439,prather2020reducing}

 \begin{equation}
 int\left\{\sum_{j=1}^{n_{sick}}0.7\times\exp\left[\frac{\left(r_i-r_j\right)^2}{2\sigma_r^2}\right]\times f\left(a_{silent},n_{out}\right) +\xi\right\}
 \end{equation}

\item For all the simulations, people are forced to maintain a distance of 3 meters from each other (social distancing). 
\end{itemize}

\section{Results}

 \begin{figure}[h!]
 \includegraphics[width=1\linewidth]{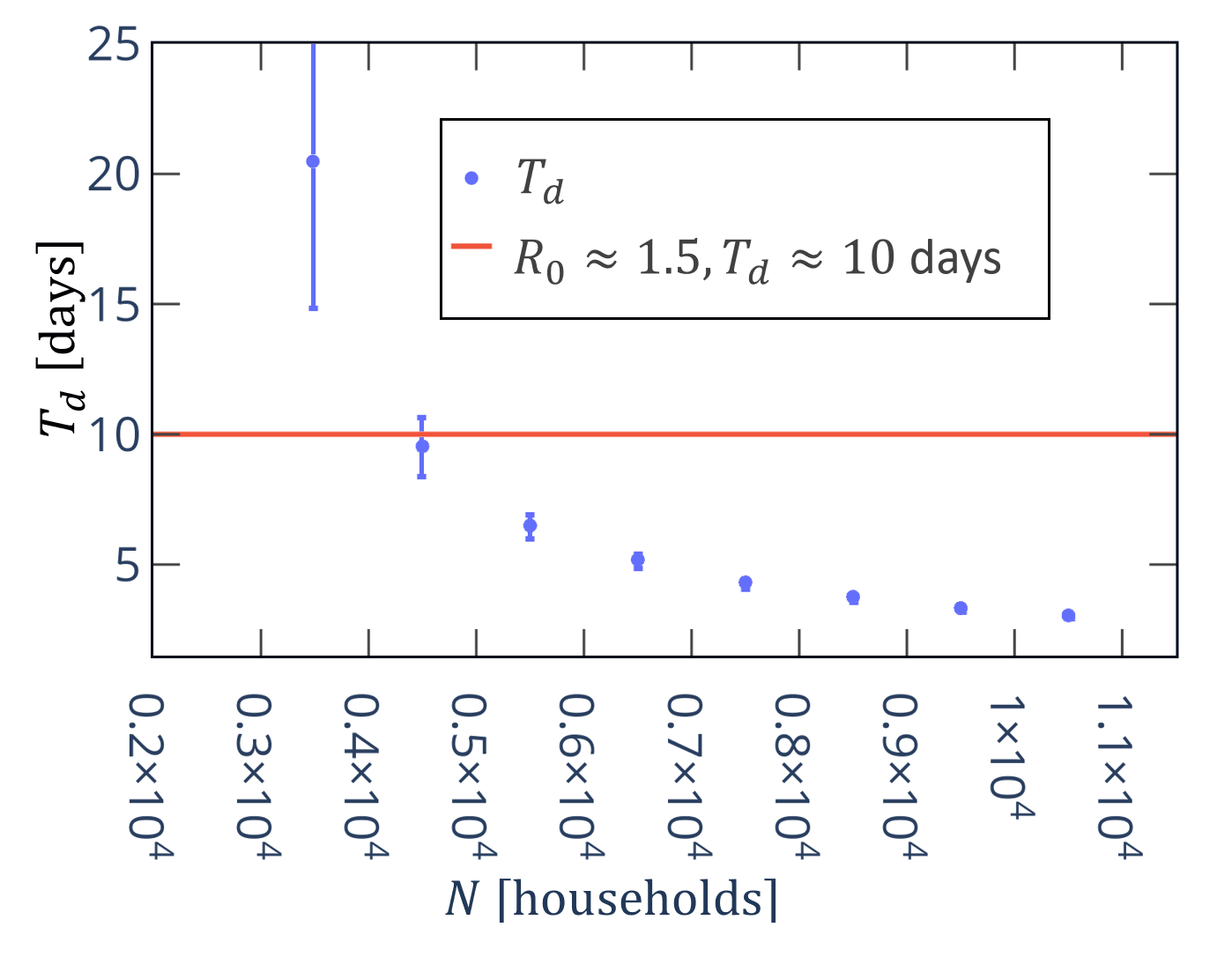}
\caption{The doubling time, $T_d$, of the percentage of active cases of the total population in the first 14 days as a function of the population density, N (dots). The solid line marks the value of $T_d$ = 10 days, corresponding to $R_t \approx 1.5$.}
\label{fig_Td}
 \end{figure}

With the continued spread of the COVID-19 virus worldwide, it appears that many countries are experiencing an increase in the rate of multiplication of patients per day at the time of outbreak stands at about doubling time, $T_d \approx 10$ days. This doubling time corresponds to $R_t \approx 1.5$, i.e., that each infection directly generates $1-2$ more infections within countermeasures like social distancing and wearing masks.

\cblack 
In our model, similar to Ref.~\cite{de2020particle}, both $R_t$ and $T_d$ are directly obtained from the simulation and not pre-assumed. We show in Fig. ~\ref{fig_Td} the doubling time, $T_d$ for the first 14 days, as a function of the effective density, $N$, under the assumption of wearing masks and social distancing. We find that a value of $T_d\approx10$ days, corresponding to $R_t=1.5$\footnote{The relation between $T_d $ and $R_t$ as shown in Fig.~\ref{fig_Td} \cblack means that attempting to describe a wide area's current situation by means of some average value of $R_t$ might be highly unappropriated. On the other hand, the spread of the disease over smaller, more homogeneous areas in which the population shares a certain degree of mobility and social behavior would be quite well described by this parameter.}, in the current model, would correspond to $N=-.45\cdot10^4$ households per square kilometer, which is significantly lower than the density used in Ref.~\cite{de2020particle} and represents the social distance actions taken in many countries.

Today, with various companies in the vaccine race, one of the questions arising these days is what vaccine rate will ensure maximum protection. In  this work, we have examined different infection scenarios under the assumption of possible re-infection for a population density of households per squared kilometer, with and without an effective vaccine. When there is also a vaccine in the following scenario, we have considered the possibility that the vaccine's effectiveness is limited in time. Note that here we present the number of active cases as a function of time (how many people are sick at any given moment) since this is the measure of the health system's capacity threshold in different countries. 

As in our previous work, the different probability densities are calculated using standard techniques, in the spirit of a kinetic Monte-Carlo simulation, for predicting the number of active Coronavirus cases in a specific area. Each case has been run 100 times \footnote{Note that each run starts with the same initial condition, only one sick person. Since the Monte Carlo algorithm is based on random numbers, we expect that every run will yield slightly different results. Repeating the simulation for 100 separated runs evaluates the algorithm's robustness. Results have been averaged and analyzed to determine the statistical error.}

\subsection{Numerical results for the case of re-infection}
\begin{figure}[h!]
 \includegraphics[width=1\linewidth]{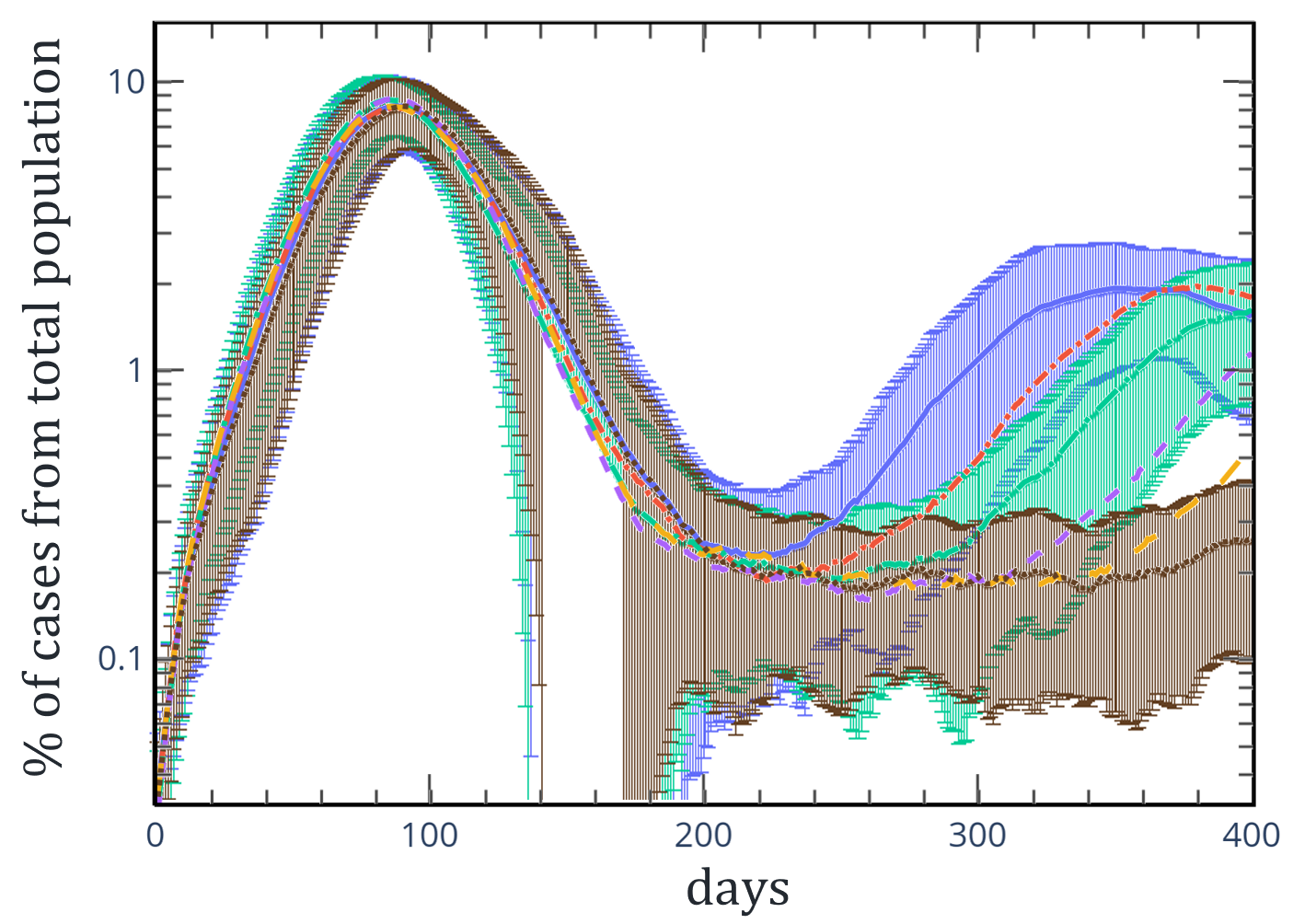}
 \caption{Numerical results for the percentage of active cases from the total population for the different re-infection cases: Solid line: 150 days; Short dashed-dotted line: 180 days; Long dashed-dotted line: 210 days; Short dashed line: 240 days; Long dashed line: 270 days; Dotted line: 300 days. Errorbars originate from the simulation’s rmsd,
computed from the 100 different samples generated to compute each curve. 
 }
 \label{fig_2I}
 \end{figure}

In this subsection, we examine how the possibility of re-infection can affect the number of active cases as a function of time. Figure~\ref{fig_2I} shows our numerical results for the different simulations for the case of $R_t=1.5$ in the first 14 days, without any further restrictions nor reliefs, which can affect $R_t$. In Fig.~\ref{fig_2I}, we show our calculations for different re-infection cases, ranging from 150 days from the time of initial infection to 300 days from the time of initial infection. From Fig.~\ref{fig_2I}, it is shown that even for low population density, re-infection can cause a second wave of infection, at an intensity that is not negligible compared to the first wave.
 
\begin{figure}[h!]
 \includegraphics[width=1\linewidth]{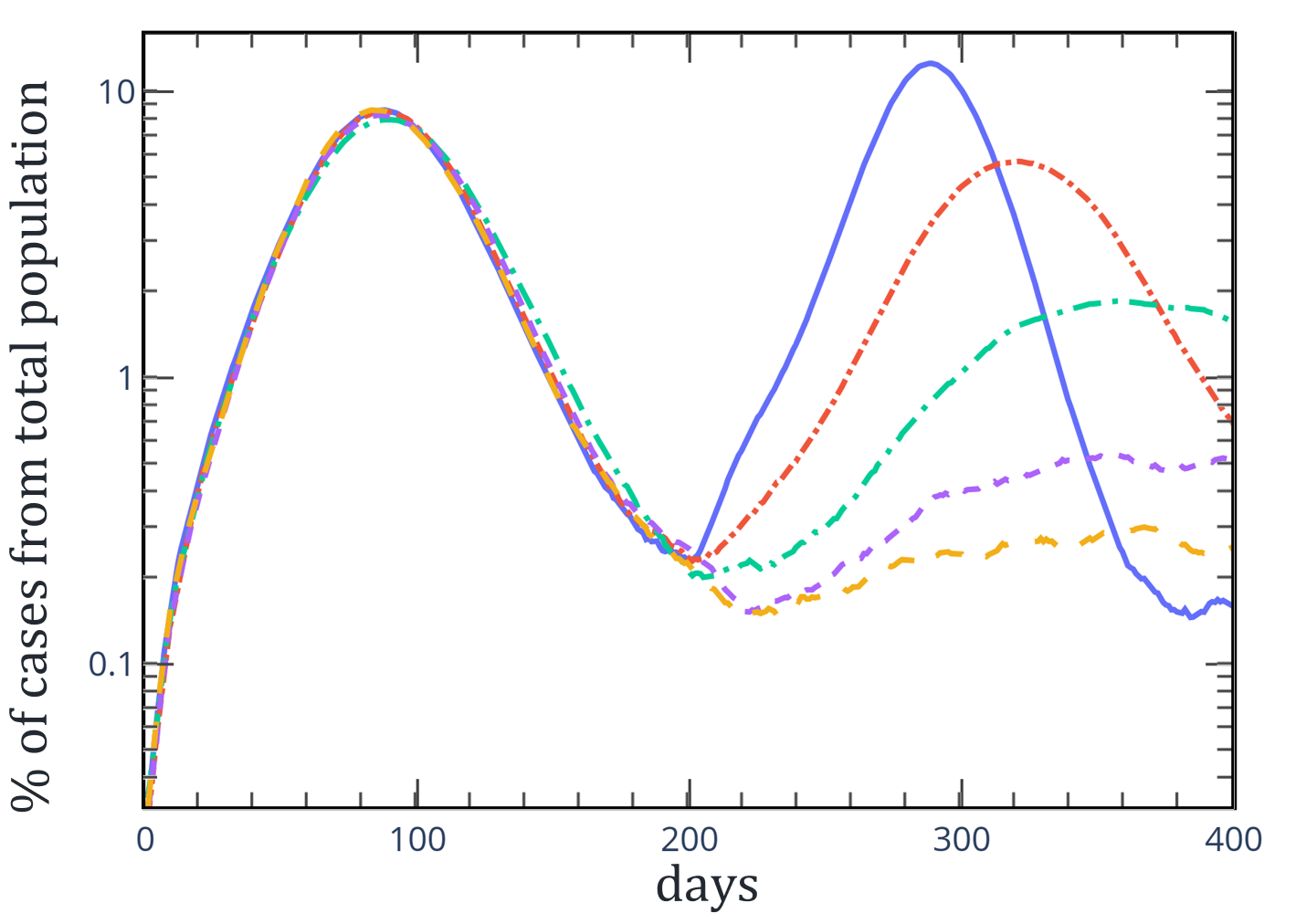}
 \caption{Numerical results for the percentage of active cases from the total population for the different surface areas (from day 200) under the assumption of the possibility of re-infection after 150 days for a population density of $N=0.45\cdot 10^4$ households. Solid line: 0.64 sq km; Short dashed-dotted: 0.8 sq km; Long dashed-dotted: 1 sq km,
 Short dashed: 1.2 sq km; Long dashed: 1.44 sq km. The uncertainty is similar to that of Fig.~\ref{fig_2I} and is omitted for improving the readability of the figure. }
 \label{fig_2I_T}
 \end{figure}

 As previously pointed out, the numerical results, presented in Fig.~\ref{fig_2I}, do not make any assumption on the doubling time, $T_d$, but only on some observed features of the disease and the population density. Hence, we examined how a temporal change of the population density, which affects $R_t$, the effective reproduction number of the virus, will be expressed in the number of cases under the assumption of re-infection. In our modal, the temporal reduction in population density is equivalent to the restrictions that are now contained worldwide due to the increase in the virus's morbidity, while a temporal increase in population density is equivalent to a release in the restrictions. Therefore, in Fig.~\ref{fig_2I_T}, we present how a temporal change of the population density from the 200$^{\text{th}}$ day (represented by a temporally change in the surface area) affects the number of active cases under the assumption of possible second infection after 150 days. Any relief will provoke a new wave of infections, even without the possibility of re-infection as long as a certain percentage of unknown infection is taken into account.
 
\begin{figure}[h!]
 \includegraphics[width=1\linewidth]{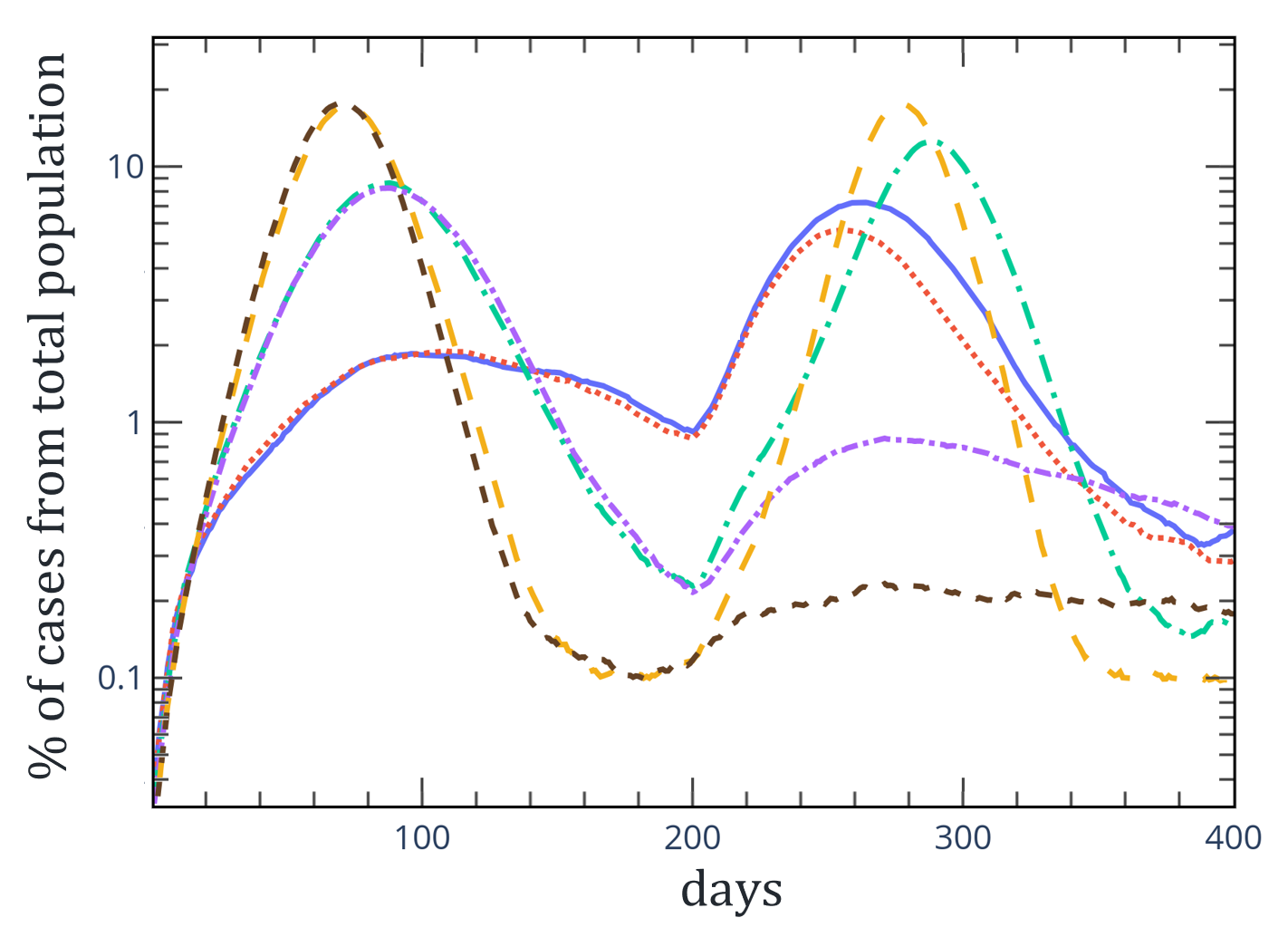}
 \caption{Numerical results for the percentage of active cases from the total population for a surface area of 0.64 sq m (from day 200) under the assumption of the possibility of re-infection after for 150 and 360 days for different population densities. Solid (Dotted) line: population density of $0.35 \cdot 10^4$ for possible re-infection after 150 (360) days; Long (Short) dashed-dotted line: population density of $0.45 \cdot 10^4$ for possible re-infection after 150 (360) days; Long (Short) dashed line: population density of $0.55 \cdot 10^4$ for possible re-infection after 150 (360) days. The uncertainty is similar to that of Fig.~\ref{fig_2I} and is omitted for improving the readability of the figure.}
 \label{fig_2I_T_180_360}
 \end{figure}

 Figure~\ref{fig_2I_T} shows that a temporal change in less than 40\% in the surface area, i.e., a change in effective population density, can cause a difference in order of magnitude in the number of infected. This result is consistent with the results from \cite{de2020particle}, which examined the effect of mobility on the coefficient of infection. Also, it is seen that increasing the surface area (which is equivalent to reducing the population density or reducing the average number of contacts per person) by 40\% can reduce the number of people infected as a function of time even without a vaccine. Hence, in Fig.~\ref{fig_2I_T_180_360} we show how a temporal increase of $R_t$ (Which in simulations is represented by
temporary change in population density), can affect another outbreak of the virus for three different population densities, also for the case that further infection can occur only after a year. 
 
Figure.~\ref{fig_2I_T_180_360} shows how increasing $R_t$ after 200 days will affect the degree of infection for different population densities. It can be seen that for the case of a low initial $R_t$ (population density of $0.3\cdot 10^4$ household per 1 sq km), increasing the population density by 1.4 times results in the second wave of infection, which is almost an order of magnitude higher than the first wave, even for a possible re-infection after a year. For a medium initial $R_t$ (population density of and 0.45 and 0.55 $\cdot 10^4$ household per 1 sq km), there is a difference between a re-infection after 150 days and after one year. In the case of prolonged immunity from further infection, we find that even for a 1.4 times increase in population density, no second wave of infection is observed due to the relatively high infection in the first wave. For a short immunity case, an increase in population density will cause a second wave similar to the first one.

Note that for all cases, a situation in which almost 10\% of the population is ill at the same time is above the capacity threshold of the health system in most of the world, such that countries will take drastic actions to reduce the rate of infection.
 
 In the next subsection, we test different vaccination situations for the case of re-infection. 

 \subsection{Numerical results for the case of effective vaccine}
 
\begin{widetext}
\begin{figure*}[t!]
 \includegraphics[width=\linewidth]{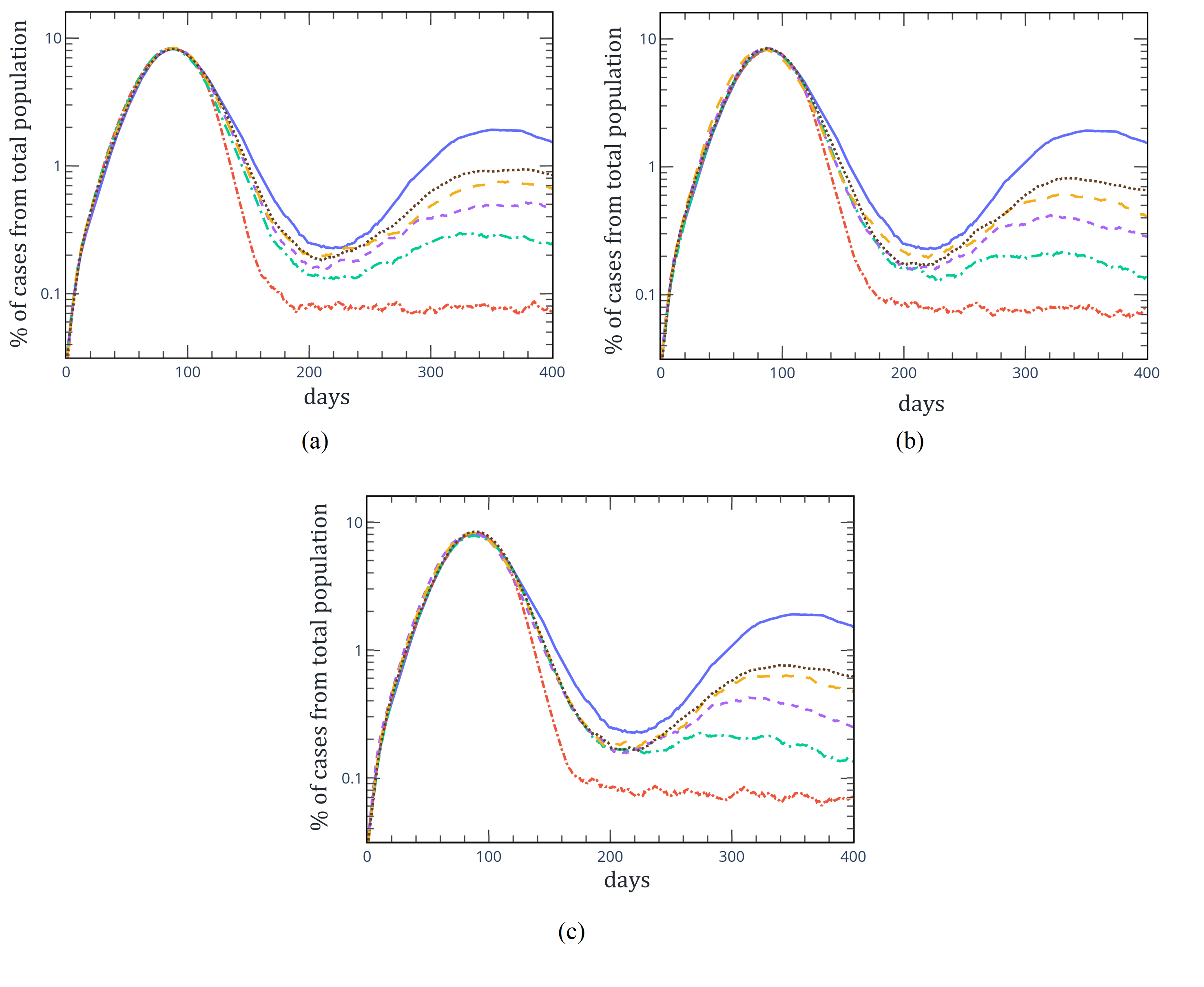}
\caption{Numerical results for the percentage of active cases from the total population for different population vaccination rates, where the vaccine is effective from the 100$^{\text{th}}$ day. Left panel (a): vaccine is effective for 180 days; Right panel (b): the vaccine is effective for 360 days; Lower panel the vaccine is effective for 720 days, (i.e., infinity, within the limitations of the simulation). For all panels - Solid line: no vaccination; Short dashed-dotted: vaccination rate of 1\%; Long dashed-dotted: vaccination rate of 0.33\%; Short dashed line: vaccination rate of 0.2\%; Long dashed line: vaccination rate of 0.15\%; Dotted line:vaccination rate of 0.1\%. The uncertainty is similar to that of Fig.~\ref{fig_2I} and is omitted for improving the readability of the figure. }
 \label{fig_2I_V}
 \vspace{-20pt}
 \end{figure*}
 \end{widetext}
 
In this subsection, we test various simulations for the case of an effective vaccine. 

In this work, we examine how the efficiency of the vaccine can reduce the number of active cases for three different cases (for all case, we assumed the re-infection without vaccination could occur after 150 days); the first, when we assume that the vaccine is valid only for half a year (180 days), i.e., after 180 days there is a chance that a person who has been vaccinated will become infected, the second, where the vaccine is efficient for a whole year and the third, where the vaccine is efficient for two years, i.e., infinity for our model. For all models, a certain percentage of the population is vaccinated starting from the 100$^{\text{th}}$ day at daily rates ranging from 1\% of the population to 0.1\% percent each day (i.e., from the 100${^\text{th}}$ days, the vaccinated populations is protected from infection).

\begin{widetext}
\begin{figure*}[t!]
 \includegraphics[width=\linewidth]{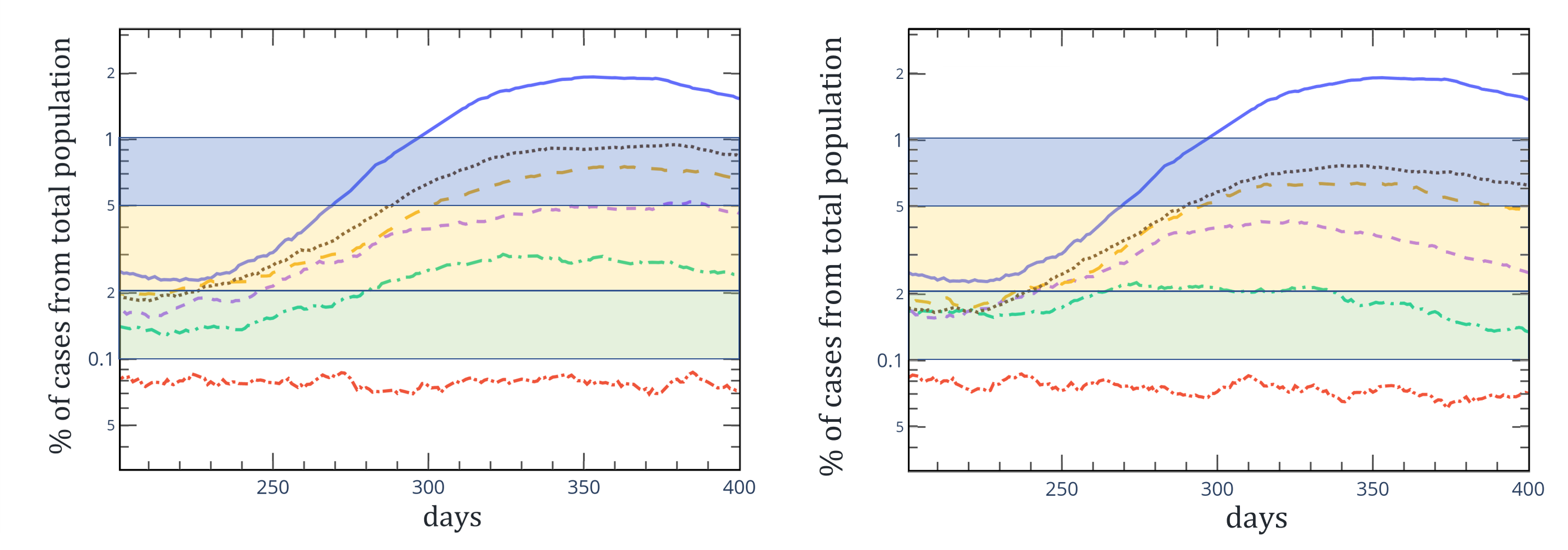}
\caption{Insert of Fig.~\ref{fig_2I_V} from the 200$^\text{th}$. Left panel (a): vaccine is effective for 180 days; Right panel (b): the vaccine is effective for 720 days, (i.e., infinity, within the limitations of the simulation). For both panels - Solid line: no vaccination; Short dashed-dotted: vaccination rate of 1\%; Long dashed-dotted: vaccination rate of 0.33\%; Short dashed line: vaccination rate of 0.2\%; Long dashed line: vaccination rate of 0.15\%; Dotted line:vaccination rate of 0.1\%. The uncertainty is similar to that of Fig.~\ref{fig_2I} and is omitted for improving the readability of the figure. }
 \label{fig_2I_V_in}
 \vspace{-20pt}
 \end{figure*}
 \end{widetext}

 Figure.~\ref{fig_2I_V} shows the number of active cases as a function of the efficacy of the vaccine and the vaccination rate (i.e., the percentage of people who can be vaccinated every day). Figure.~\ref{fig_2I_V} indicates that the impact of the vaccination rate on the daily number of patients is much higher than its efficiency. From Fig.~\ref{fig_2I_V_in}, we see that a vaccination rate of 0.1\% of the population per day (which is equivalent to$\sim$300,000 vaccine doses daily in the United States for at least a year) will not significantly reduce morbidity even for an infinite efficiency. Figure.~\ref{fig_2I_V_in} is an insert of Fig.~\ref{fig_2I_V} from the 200$^{\text{th}}$ day, where different percentages are marked for convenience. From Fig.~\ref{fig_2I_V_in} it easy to see that when a $\sim1\%$ of the population can be vaccinated daily, the effectiveness of the vaccine does not affect morbidity since the rate of immunization is higher than the rate of decline in immune protection even for the efficacy of only 180 days.
On the other hand, when the rate of protection decreases faster than the rate of vaccination of the population, there is a difference between the efficiencies of 180 days and 720 days (i.e., infinity). For example, for a 720-day immunization efficacy, a vaccination rate of 0.2\% of the population per day would result in similar immunization rates of 0.33\% of the population per day, for a vaccination efficacy of only 180 days, where for both cases, the maximal percentage of infected people is $\sim 0.5\%$. Therefore, as long as the effectiveness of the vaccine is unknown and the entire population cannot be massively vaccinated, it will be necessary to maintain the social restrictions to prevent outbreaks of the pandemic (i.e., a vaccination rate of 0.14\% means that about one million people in Europe will be vaccinated daily).
 
 Also, by comparing to Fig.~\ref{fig_2I_T}, we find that reducing the effective population density by 20\% (which can be achieved by enforcing various social constraints) leads to a vaccine-like effect of a daily vaccination rate of 0.2\% of the entire population. Hence, social distancing and wearing masks will have to be maintained even after vaccinating part of the population to reduce the $R_t$ as much as possible and to prevent another outbreak of the pandemic.
 Hence, it of interest to examine the effect of temporary changes in $R_t$, in the case of an effective vaccine (720 days) with a vaccination rate of approximately 0.2\% of the population each day. We assume that naturally, vaccination of the population will lead to a loosening of the strictness of the restrictions, which, as a result, will affect $R_t$.
 \begin{figure}[h!]
 \includegraphics[width=1\linewidth]{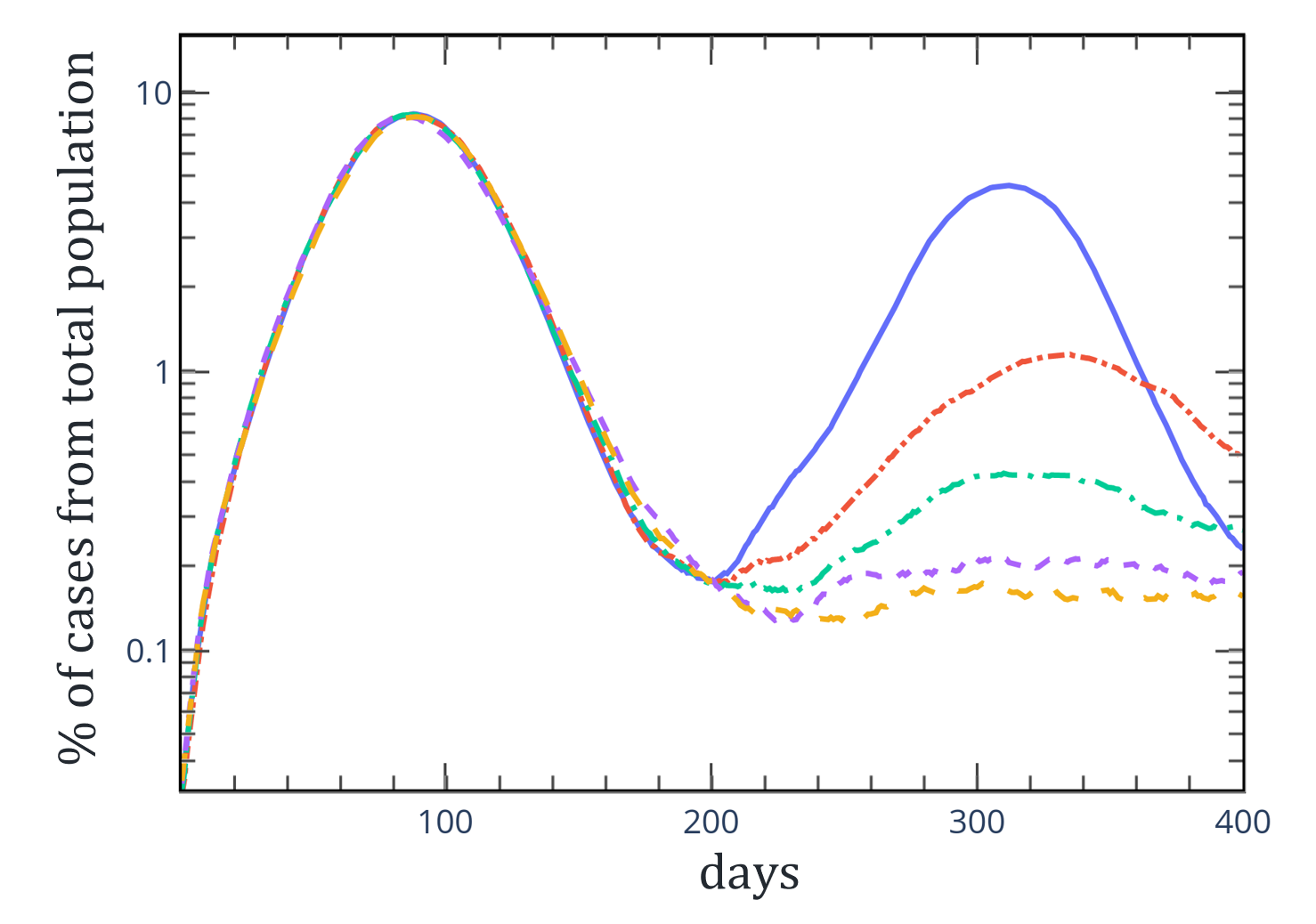}
 \caption{Numerical results for the percentage of active cases from the total population for different surface areas (from day 200) under the assumption of the possibility of re-infection after 150 days for a population density of $N=0.45\cdot 10^4$ households, with a vaccination rate of 0.2\% per day. Solid line:  0.64 sq km; Short dashed-dotted line: 0.8 sq km; Long dashed-dotted line: 1 sq km;
 Short dashed line: 1.2 sq km; Long dashed- line: 1.44 sq km. The uncertainty is similar to that of Fig.~\ref{fig_2I} and is omitted for improving the readability of the figure. }
 \label{fig_2I_V_T_720}
 \end{figure}
 
 Figure.~\ref{fig_2I_V_T_720} indicates that for a vaccination rate of 0.2 \%, an uncontrolled release of the restrictions will lead to an increase in $R_t$ and, and as a result, to a further outbreak of the pandemic. For example, increasing the population density by about 20\% will result in a maximum active percentage of patients of 2 \% . On the other hand, vaccinating the population, which will be accompanied by restrictions that parallel the reduction of the population density by ~ 20 \%, will result in a maximum active percentage of patients of 0.2 \%. Hence, during the immunization period, the public must observe the restrictions and wear masks, and that masks have to be worn until a significant percentage of the population was vaccinated. 
\cblack

 \section{Conclusions}
 In this paper, we presented a kinetic Monte-Carlo algorithm for modeling different scenarios of spreading the COVID-19 under the assumption of possible re-infection and an efficient vaccine. The model's main advantage is its extreme flexibility, such that we can test many infection scenarios with a minimal computational cost. 

In this work, we studied how the possibility of re-infection affects the epidemic outbreak for different scenarios.
Also, we examined how a temporary change in $R_t$, the effective reproduction rate of the virus, has an influence on the re-emergence of the pandemic when we assume the possibility of re-infection and infection from unknown sources, and how
an effective vaccine would affect an infection recurrence in this situation. We found that even with a relatively low effective reproduction rate of the virus, i.e., $R_t\approx1.5$, (which is a result of social distancing and local restrictions), a significant percentage of the population should be vaccinated to prevent the outbreak of a virus.

Hence, in principle, our model might be used as a tool for decision-making concerning the rate of vaccination and for deciding which populations should be vaccinated first for a limited amount of vaccines, depending on the healthcare system capacity threshold in each country.

Nevertheless, our results show that as long as a significant percentage of the population is not vaccinated, social distance regulations will have to be enforced, and people will continue to wear masks to reduce $R_t$ as much as possible to prevent an outbreak of the pandemic.

\begin{acknowledgments}

We thank R. Milo and his group at Weizmann Institute of Science for sharing their epidemiological data and for fruitful discussions. We thank 
G. Haimovich at Weizmann Institute of Science for fruitful discussions. \end{acknowledgments}

The data that support the findings of this study are available from the corresponding author upon reasonable request.

\bibliography{references}

\begin{thebibliography}{27}%
\makeatletter
\providecommand \@ifxundefined [1]{%
 \@ifx{#1\undefined}
}%
\providecommand \@ifnum [1]{%
 \ifnum #1\expandafter \@firstoftwo
 \else \expandafter \@secondoftwo
 \fi
}%
\providecommand \@ifx [1]{%
 \ifx #1\expandafter \@firstoftwo
 \else \expandafter \@secondoftwo
 \fi
}%
\providecommand \natexlab [1]{#1}%
\providecommand \enquote  [1]{``#1''}%
\providecommand \bibnamefont  [1]{#1}%
\providecommand \bibfnamefont [1]{#1}%
\providecommand \citenamefont [1]{#1}%
\providecommand \href@noop [0]{\@secondoftwo}%
\providecommand \href [0]{\begingroup \@sanitize@url \@href}%
\providecommand \@href[1]{\@@startlink{#1}\@@href}%
\providecommand \@@href[1]{\endgroup#1\@@endlink}%
\providecommand \@sanitize@url [0]{\catcode `\\12\catcode `\$12\catcode
  `\&12\catcode `\#12\catcode `\^12\catcode `\_12\catcode `\%12\relax}%
\providecommand \@@startlink[1]{}%
\providecommand \@@endlink[0]{}%
\providecommand \url  [0]{\begingroup\@sanitize@url \@url }%
\providecommand \@url [1]{\endgroup\@href {#1}{\urlprefix }}%
\providecommand \urlprefix  [0]{URL }%
\providecommand \Eprint [0]{\href }%
\providecommand \doibase [0]{http://dx.doi.org/}%
\providecommand \selectlanguage [0]{\@gobble}%
\providecommand \bibinfo  [0]{\@secondoftwo}%
\providecommand \bibfield  [0]{\@secondoftwo}%
\providecommand \translation [1]{[#1]}%
\providecommand \BibitemOpen [0]{}%
\providecommand \bibitemStop [0]{}%
\providecommand \bibitemNoStop [0]{.\EOS\space}%
\providecommand \EOS [0]{\spacefactor3000\relax}%
\providecommand \BibitemShut  [1]{\csname bibitem#1\endcsname}%
\let\auto@bib@innerbib\@empty
%</preamble>
\bibitem [{pfi()}]{pfizer_2020}%
  \BibitemOpen
  \bibfield  {title} {\enquote {\bibinfo {title} {Pfizer and biontech announce
  vaccine candidate against covid-19 achieved success in first interim analysis
  from phase 3 study},}\ }\href@noop {} {\bibinfo  {journal}
  {https://www.pfizer.com/news/press-release/press-release-detail/pfizer-and-biontech-announce-vaccine-candidate-against}\
  }\BibitemShut {NoStop}%
\bibitem [{mod()}]{modernas_2020}%
  \BibitemOpen
\bibfield  {journal} {  }\bibfield  {title} {\enquote {\bibinfo {title}
  {Moderna’s covid-19 vaccine candidate meets its primary efficacy endpoint
  in the first interim analysis of the phase 3 cove study},}\ }\href@noop {}
  {\bibinfo  {journal}
  {https://investors.modernatx.com/news-releases/news-release-details/modernas-covid-19-vaccine-candidate-meets-its-primary-efficacy}\
  }\BibitemShut {NoStop}%
\bibitem [{\citenamefont {Gousseff}\ \emph {et~al.}(2020)\citenamefont
  {Gousseff}, \citenamefont {Penot}, \citenamefont {Gallay}, \citenamefont
  {Batisse}, \citenamefont {Benech}, \citenamefont {Bouiller}, \citenamefont
  {Collarino}, \citenamefont {Conrad}, \citenamefont {Slama}, \citenamefont
  {Joseph}, \citenamefont {Lemaignen}, \citenamefont {Lescure}, \citenamefont
  {Levy}, \citenamefont {Mahevas}, \citenamefont {Pozzetto}, \citenamefont
  {Vignier}, \citenamefont {Wyplosz}, \citenamefont {Salmon}, \citenamefont
  {Goehringer},\ and\ \citenamefont {Botelho-Nevers}}]{GOUSSEFF2020816}%
  \BibitemOpen
\bibfield  {journal} {  }\bibfield  {author} {\bibinfo {author} {\bibfnamefont
  {M.}~\bibnamefont {Gousseff}}, \bibinfo {author} {\bibfnamefont
  {P.}~\bibnamefont {Penot}}, \bibinfo {author} {\bibfnamefont
  {L.}~\bibnamefont {Gallay}}, \bibinfo {author} {\bibfnamefont
  {D.}~\bibnamefont {Batisse}}, \bibinfo {author} {\bibfnamefont
  {N.}~\bibnamefont {Benech}}, \bibinfo {author} {\bibfnamefont
  {K.}~\bibnamefont {Bouiller}}, \bibinfo {author} {\bibfnamefont
  {R.}~\bibnamefont {Collarino}}, \bibinfo {author} {\bibfnamefont
  {A.}~\bibnamefont {Conrad}}, \bibinfo {author} {\bibfnamefont
  {D.}~\bibnamefont {Slama}}, \bibinfo {author} {\bibfnamefont
  {C.}~\bibnamefont {Joseph}}, \bibinfo {author} {\bibfnamefont
  {A.}~\bibnamefont {Lemaignen}}, \bibinfo {author} {\bibfnamefont {F.-X.}\
  \bibnamefont {Lescure}}, \bibinfo {author} {\bibfnamefont {B.}~\bibnamefont
  {Levy}}, \bibinfo {author} {\bibfnamefont {M.}~\bibnamefont {Mahevas}},
  \bibinfo {author} {\bibfnamefont {B.}~\bibnamefont {Pozzetto}}, \bibinfo
  {author} {\bibfnamefont {N.}~\bibnamefont {Vignier}}, \bibinfo {author}
  {\bibfnamefont {B.}~\bibnamefont {Wyplosz}}, \bibinfo {author} {\bibfnamefont
  {D.}~\bibnamefont {Salmon}}, \bibinfo {author} {\bibfnamefont
  {F.}~\bibnamefont {Goehringer}}, \ and\ \bibinfo {author} {\bibfnamefont
  {E.}~\bibnamefont {Botelho-Nevers}},\ }\bibfield  {title} {\enquote {\bibinfo
  {title} {Clinical recurrences of covid-19 symptoms after recovery: Viral
  relapse, reinfection or inflammatory rebound?}}\ }\href {\doibase
  https://doi.org/10.1016/j.jinf.2020.06.073} {\bibfield  {journal} {\bibinfo
  {journal} {Journal of Infection}\ }\textbf {\bibinfo {volume} {81}},\
  \bibinfo {pages} {816 -- 846} (\bibinfo {year} {2020})}\BibitemShut {NoStop}%
\bibitem [{\citenamefont {Duggan}\ \emph {et~al.}(2020)\citenamefont {Duggan},
  \citenamefont {Ludy}, \citenamefont {Shannon}, \citenamefont {Reisner},\ and\
  \citenamefont {Wilcox}}]{duggan2020case}%
  \BibitemOpen
  \bibfield  {author} {\bibinfo {author} {\bibfnamefont {N.~M.}\ \bibnamefont
  {Duggan}}, \bibinfo {author} {\bibfnamefont {S.~M.}\ \bibnamefont {Ludy}},
  \bibinfo {author} {\bibfnamefont {B.~C.}\ \bibnamefont {Shannon}}, \bibinfo
  {author} {\bibfnamefont {A.~T.}\ \bibnamefont {Reisner}}, \ and\ \bibinfo
  {author} {\bibfnamefont {S.~R.}\ \bibnamefont {Wilcox}},\ }\bibfield  {title}
  {\enquote {\bibinfo {title} {A case report of possible novel coronavirus 2019
  reinfection},}\ }\href@noop {} {\bibfield  {journal} {\bibinfo  {journal}
  {The American journal of emergency medicine}\ } (\bibinfo {year}
  {2020})}\BibitemShut {NoStop}%
\bibitem [{\citenamefont {Alizargar}(2020)}]{alizargar2020risk}%
  \BibitemOpen
  \bibfield  {author} {\bibinfo {author} {\bibfnamefont {J.}~\bibnamefont
  {Alizargar}},\ }\bibfield  {title} {\enquote {\bibinfo {title} {Risk of
  reactivation or reinfection of novel coronavirus (covid-19)},}\ }\href@noop
  {} {\bibfield  {journal} {\bibinfo  {journal} {Journal of the Formosan
  Medical Association}\ } (\bibinfo {year} {2020})}\BibitemShut {NoStop}%
\bibitem [{\citenamefont {Edridge}\ \emph {et~al.}(2020)\citenamefont
  {Edridge}, \citenamefont {Kaczorowska}, \citenamefont {Hoste}, \citenamefont
  {Bakker}, \citenamefont {Klein}, \citenamefont {Loens}, \citenamefont
  {Jebbink}, \citenamefont {Matser}, \citenamefont {Kinsella}, \citenamefont
  {Rueda} \emph {et~al.}}]{edridge2020seasonal}%
  \BibitemOpen
  \bibfield  {author} {\bibinfo {author} {\bibfnamefont {A.~W.}\ \bibnamefont
  {Edridge}}, \bibinfo {author} {\bibfnamefont {J.}~\bibnamefont
  {Kaczorowska}}, \bibinfo {author} {\bibfnamefont {A.~C.}\ \bibnamefont
  {Hoste}}, \bibinfo {author} {\bibfnamefont {M.}~\bibnamefont {Bakker}},
  \bibinfo {author} {\bibfnamefont {M.}~\bibnamefont {Klein}}, \bibinfo
  {author} {\bibfnamefont {K.}~\bibnamefont {Loens}}, \bibinfo {author}
  {\bibfnamefont {M.~F.}\ \bibnamefont {Jebbink}}, \bibinfo {author}
  {\bibfnamefont {A.}~\bibnamefont {Matser}}, \bibinfo {author} {\bibfnamefont
  {C.~M.}\ \bibnamefont {Kinsella}}, \bibinfo {author} {\bibfnamefont
  {P.}~\bibnamefont {Rueda}},  \emph {et~al.},\ }\bibfield  {title} {\enquote
  {\bibinfo {title} {Seasonal coronavirus protective immunity is
  short-lasting},}\ }\href@noop {} {\bibfield  {journal} {\bibinfo  {journal}
  {Nature medicine}\ ,\ \bibinfo {pages} {1--3}} (\bibinfo {year}
  {2020})}\BibitemShut {NoStop}%
\bibitem [{\citenamefont {Landi}\ \emph {et~al.}(2020)\citenamefont {Landi},
  \citenamefont {Carf{\`\i}}, \citenamefont {Benvenuto}, \citenamefont
  {Brandi}, \citenamefont {Ciciarello}, \citenamefont {Monaco}, \citenamefont
  {Martone}, \citenamefont {Napolitano}, \citenamefont {Pagano}, \citenamefont
  {Paglionico} \emph {et~al.}}]{landi2020predictive}%
  \BibitemOpen
  \bibfield  {author} {\bibinfo {author} {\bibfnamefont {F.}~\bibnamefont
  {Landi}}, \bibinfo {author} {\bibfnamefont {A.}~\bibnamefont {Carf{\`\i}}},
  \bibinfo {author} {\bibfnamefont {F.}~\bibnamefont {Benvenuto}}, \bibinfo
  {author} {\bibfnamefont {V.}~\bibnamefont {Brandi}}, \bibinfo {author}
  {\bibfnamefont {F.}~\bibnamefont {Ciciarello}}, \bibinfo {author}
  {\bibfnamefont {M.~R.~L.}\ \bibnamefont {Monaco}}, \bibinfo {author}
  {\bibfnamefont {A.~M.}\ \bibnamefont {Martone}}, \bibinfo {author}
  {\bibfnamefont {C.}~\bibnamefont {Napolitano}}, \bibinfo {author}
  {\bibfnamefont {F.}~\bibnamefont {Pagano}}, \bibinfo {author} {\bibfnamefont
  {A.}~\bibnamefont {Paglionico}},  \emph {et~al.},\ }\bibfield  {title}
  {\enquote {\bibinfo {title} {Predictive factors for a new positive
  nasopharyngeal swab among patients recovered from covid-19},}\ }\href@noop {}
  {\bibfield  {journal} {\bibinfo  {journal} {American Journal of Preventive
  Medicine}\ } (\bibinfo {year} {2020})}\BibitemShut {NoStop}%
\bibitem [{sci()}]{science_2020}%
  \BibitemOpen
  \bibfield  {title} {\enquote {\bibinfo {title} {More people are getting
  covid-19 twice, suggesting immunity wanes quickly in some},}\ }\href@noop {}
  {\bibinfo  {journal}
  {https://www.sciencemag.org/news/2020/11/more-people-are-getting-covid-19-twice-suggesting-immunity-wanes-quickly-some}\
  }\BibitemShut {NoStop}%
\bibitem [{\citenamefont {De-Leon}\ and\ \citenamefont
  {Pederiva}(2020)}]{de2020particle}%
  \BibitemOpen
\bibfield  {journal} {  }\bibfield  {author} {\bibinfo {author} {\bibfnamefont
  {H.}~\bibnamefont {De-Leon}}\ and\ \bibinfo {author} {\bibfnamefont
  {F.}~\bibnamefont {Pederiva}},\ }\bibfield  {title} {\enquote {\bibinfo
  {title} {Particle modeling of the spreading of coronavirus disease
  (covid-19)},}\ }\href {\doibase 10.1063/5.0020565} {\bibfield  {journal}
  {\bibinfo  {journal} {Physics of Fluids}\ }\textbf {\bibinfo {volume} {32}},\
  \bibinfo {pages} {087113} (\bibinfo {year} {2020})},\ \Eprint
  {http://arxiv.org/abs/https://doi.org/10.1063/5.0020565}
  {https://doi.org/10.1063/5.0020565} \BibitemShut {NoStop}%
\bibitem [{Note1()}]{Note1}%
  \BibitemOpen
  \bibinfo {note} {Note that the onset of the epidemic was characterized by the
  parameter $R_0$, the basic reproducing number. Now, in many countries, social
  distance measures are taken that affect the spread of the epidemic. Hence,
  the calculations in this work were done under the assumption that an effort
  is made to reduce $R_t$ by various constraints.}\BibitemShut {Stop}%
\bibitem [{\citenamefont {Bar-On}\ \emph {et~al.}(2020)\citenamefont {Bar-On},
  \citenamefont {Flamholz}, \citenamefont {Phillips},\ and\ \citenamefont
  {Milo}}]{10.7554/eLife.57309}%
  \BibitemOpen
  \bibfield  {author} {\bibinfo {author} {\bibfnamefont {Y.~M.}\ \bibnamefont
  {Bar-On}}, \bibinfo {author} {\bibfnamefont {A.}~\bibnamefont {Flamholz}},
  \bibinfo {author} {\bibfnamefont {R.}~\bibnamefont {Phillips}}, \ and\
  \bibinfo {author} {\bibfnamefont {R.}~\bibnamefont {Milo}},\ }\bibfield
  {title} {\enquote {\bibinfo {title} {Sars-cov-2 (covid-19) by the numbers},}\
  }\href {\doibase 10.7554/eLife.57309} {\bibfield  {journal} {\bibinfo
  {journal} {eLife}\ }\textbf {\bibinfo {volume} {9}},\ \bibinfo {pages}
  {e57309} (\bibinfo {year} {2020})}\BibitemShut {NoStop}%
\bibitem [{\citenamefont {Bhardwaj}\ and\ \citenamefont
  {Agrawal}(2020)}]{doi:10.1063/5.0012009}%
  \BibitemOpen
  \bibfield  {author} {\bibinfo {author} {\bibfnamefont {R.}~\bibnamefont
  {Bhardwaj}}\ and\ \bibinfo {author} {\bibfnamefont {A.}~\bibnamefont
  {Agrawal}},\ }\bibfield  {title} {\enquote {\bibinfo {title} {Likelihood of
  survival of coronavirus in a respiratory droplet deposited on a solid
  surface},}\ }\href {\doibase 10.1063/5.0012009} {\bibfield  {journal}
  {\bibinfo  {journal} {Physics of Fluids}\ }\textbf {\bibinfo {volume} {32}},\
  \bibinfo {pages} {061704} (\bibinfo {year} {2020})},\ \Eprint
  {http://arxiv.org/abs/https://doi.org/10.1063/5.0012009}
  {https://doi.org/10.1063/5.0012009} \BibitemShut {NoStop}%
\bibitem [{\citenamefont {Dbouk}\ and\ \citenamefont
  {Drikakis}(2020{\natexlab{a}})}]{doi:10.1063/5.0011960}%
  \BibitemOpen
  \bibfield  {author} {\bibinfo {author} {\bibfnamefont {T.}~\bibnamefont
  {Dbouk}}\ and\ \bibinfo {author} {\bibfnamefont {D.}~\bibnamefont
  {Drikakis}},\ }\bibfield  {title} {\enquote {\bibinfo {title} {On coughing
  and airborne droplet transmission to humans},}\ }\href {\doibase
  10.1063/5.0011960} {\bibfield  {journal} {\bibinfo  {journal} {Physics of
  Fluids}\ }\textbf {\bibinfo {volume} {32}},\ \bibinfo {pages} {053310}
  (\bibinfo {year} {2020}{\natexlab{a}})},\ \Eprint
  {http://arxiv.org/abs/https://doi.org/10.1063/5.0011960}
  {https://doi.org/10.1063/5.0011960} \BibitemShut {NoStop}%
\bibitem [{Note2()}]{Note2}%
  \BibitemOpen
  \bibinfo {note} {We are aware that this model cannot take into account every
  single spreading event, but since such events affect the initial $T_d$,
  (which is a function of the population density), and since the infection
  process is random, we expect that the existence of such events will be
  reflected in the numerical results. Also, in contrast to real-life, here,
  there is no time gap between the infection and being tested positive for
  Coronavirus. Therefore, an immediate decrease in the rate of infection
  resulting from lockdown is expected in the model in contrast to the real
  data. \cite {endcoronavirus}.}\BibitemShut {Stop}%
\bibitem [{Note3()}]{Note3}%
  \BibitemOpen
  \bibinfo {note} {The application of periodic boundary conditions means that
  we have an infinite number of identical systems; each system is a replica of
  the others. I.e., if a person leaves the simulation surface on one side, an
  equal person will enter the surface from the other side.}\BibitemShut {Stop}%
\bibitem [{\citenamefont {Li}\ \emph {et~al.}(2020)\citenamefont {Li},
  \citenamefont {Zhang}, \citenamefont {Lu}, \citenamefont {Liu}, \citenamefont
  {Chang}, \citenamefont {Cao}, \citenamefont {Liu}, \citenamefont {Zhang},
  \citenamefont {Ling}, \citenamefont {Tao},\ and\ \citenamefont
  {Chen}}]{10.1093/cid/ciaa450}%
  \BibitemOpen
  \bibfield  {author} {\bibinfo {author} {\bibfnamefont {W.}~\bibnamefont
  {Li}}, \bibinfo {author} {\bibfnamefont {B.}~\bibnamefont {Zhang}}, \bibinfo
  {author} {\bibfnamefont {J.}~\bibnamefont {Lu}}, \bibinfo {author}
  {\bibfnamefont {S.}~\bibnamefont {Liu}}, \bibinfo {author} {\bibfnamefont
  {Z.}~\bibnamefont {Chang}}, \bibinfo {author} {\bibfnamefont
  {P.}~\bibnamefont {Cao}}, \bibinfo {author} {\bibfnamefont {X.}~\bibnamefont
  {Liu}}, \bibinfo {author} {\bibfnamefont {P.}~\bibnamefont {Zhang}}, \bibinfo
  {author} {\bibfnamefont {Y.}~\bibnamefont {Ling}}, \bibinfo {author}
  {\bibfnamefont {K.}~\bibnamefont {Tao}}, \ and\ \bibinfo {author}
  {\bibfnamefont {J.}~\bibnamefont {Chen}},\ }\bibfield  {title} {\enquote
  {\bibinfo {title} {{The characteristics of household transmission of
  COVID-19}},}\ }\href {\doibase 10.1093/cid/ciaa450} {\bibfield  {journal}
  {\bibinfo  {journal} {Clinical Infectious Diseases}\ } (\bibinfo {year}
  {2020}),\ 10.1093/cid/ciaa450},\ \bibinfo {note} {ciaa450},\ \Eprint
  {http://arxiv.org/abs/https://academic.oup.com/cid/advance-article-pdf/doi/10.1093/cid/ciaa450/33097408/ciaa450.pdf}
  {https://academic.oup.com/cid/advance-article-pdf/doi/10.1093/cid/ciaa450/33097408/ciaa450.pdf}
  \BibitemShut {NoStop}%
\bibitem [{Note4()}]{Note4}%
  \BibitemOpen
  \bibinfo {note} {Since the infection probability is a function of the
  absolute value of the distance between two people and the standard deviation,
  several distributions such as a Gaussian distribution and a Lorentzian
  distribution could serve for modeling the infection probability under the
  assumption that the simulation's dynamics dependents on $\sigma _r$ and not
  on the distribution's tail. The choice of Gaussian distribution was since
  this is the typical distribution for thermal systems approaches for
  equilibrium, e.g., Maxwell Boltzmann distribution.}\BibitemShut {Stop}%
\bibitem [{\citenamefont {Jing}\ \emph {et~al.}(2020)\citenamefont {Jing},
  \citenamefont {Liu}, \citenamefont {Yuan}, \citenamefont {Zhang},
  \citenamefont {Zhang}, \citenamefont {Dean}, \citenamefont {Luo},
  \citenamefont {Ma}, \citenamefont {Longini}, \citenamefont {Kenah} \emph
  {et~al.}}]{jing2020household}%
  \BibitemOpen
  \bibfield  {author} {\bibinfo {author} {\bibfnamefont {Q.-L.}\ \bibnamefont
  {Jing}}, \bibinfo {author} {\bibfnamefont {M.-J.}\ \bibnamefont {Liu}},
  \bibinfo {author} {\bibfnamefont {J.}~\bibnamefont {Yuan}}, \bibinfo {author}
  {\bibfnamefont {Z.-B.}\ \bibnamefont {Zhang}}, \bibinfo {author}
  {\bibfnamefont {A.-R.}\ \bibnamefont {Zhang}}, \bibinfo {author}
  {\bibfnamefont {N.~E.}\ \bibnamefont {Dean}}, \bibinfo {author}
  {\bibfnamefont {L.}~\bibnamefont {Luo}}, \bibinfo {author} {\bibfnamefont
  {M.-M.}\ \bibnamefont {Ma}}, \bibinfo {author} {\bibfnamefont
  {I.}~\bibnamefont {Longini}}, \bibinfo {author} {\bibfnamefont
  {E.}~\bibnamefont {Kenah}},  \emph {et~al.},\ }\bibfield  {title} {\enquote
  {\bibinfo {title} {Household secondary attack rate of covid-19 and associated
  determinants},}\ }\href@noop {} {\bibfield  {journal} {\bibinfo  {journal}
  {medRxiv}\ } (\bibinfo {year} {2020})}\BibitemShut {NoStop}%
\bibitem [{\citenamefont {Dbouk}\ and\ \citenamefont
  {Drikakis}(2020{\natexlab{b}})}]{doi:10.1063/5.0015044}%
  \BibitemOpen
  \bibfield  {author} {\bibinfo {author} {\bibfnamefont {T.}~\bibnamefont
  {Dbouk}}\ and\ \bibinfo {author} {\bibfnamefont {D.}~\bibnamefont
  {Drikakis}},\ }\bibfield  {title} {\enquote {\bibinfo {title} {On respiratory
  droplets and face masks},}\ }\href {\doibase 10.1063/5.0015044} {\bibfield
  {journal} {\bibinfo  {journal} {Physics of Fluids}\ }\textbf {\bibinfo
  {volume} {32}},\ \bibinfo {pages} {063303} (\bibinfo {year}
  {2020}{\natexlab{b}})},\ \Eprint
  {http://arxiv.org/abs/https://doi.org/10.1063/5.0015044}
  {https://doi.org/10.1063/5.0015044} \BibitemShut {NoStop}%
\bibitem [{19_(2020)}]{19_2020}%
  \BibitemOpen
  \bibfield  {title} {\enquote {\bibinfo {title} {Hku hamster research shows
  masks effective in preventing covid-19 transmission - fight covid},}\
  }\href@noop {} {\bibfield  {journal} {\bibinfo  {journal}
  {https://medicalxpress.com/news/2020-05-hamster-masks-coronavirus-scientists.html/}\
  } (\bibinfo {year} {2020})}\BibitemShut {NoStop}%
\bibitem [{\citenamefont {Robinson}\ \emph {et~al.}(2020)\citenamefont
  {Robinson}, \citenamefont {de~Anda}, \citenamefont {Moore}, \citenamefont
  {Reid}, \citenamefont {Sear},\ and\ \citenamefont
  {Royall}}]{robinson2020efficacy}%
  \BibitemOpen
  \bibfield  {author} {\bibinfo {author} {\bibfnamefont {J.~F.}\ \bibnamefont
  {Robinson}}, \bibinfo {author} {\bibfnamefont {I.~R.}\ \bibnamefont
  {de~Anda}}, \bibinfo {author} {\bibfnamefont {F.}~\bibnamefont {Moore}},
  \bibinfo {author} {\bibfnamefont {J.~P.}\ \bibnamefont {Reid}}, \bibinfo
  {author} {\bibfnamefont {R.~P.}\ \bibnamefont {Sear}}, \ and\ \bibinfo
  {author} {\bibfnamefont {C.~P.}\ \bibnamefont {Royall}},\ }\bibfield  {title}
  {\enquote {\bibinfo {title} {Efficacy of face coverings in reducing
  transmission of covid-19: calculations based on models of droplet capture},}\
  }\href@noop {} {\bibfield  {journal} {\bibinfo  {journal} {arXiv preprint
  arXiv:2008.04995}\ } (\bibinfo {year} {2020})}\BibitemShut {NoStop}%
\bibitem [{\citenamefont {Eikenberry}\ \emph {et~al.}(2020)\citenamefont
  {Eikenberry}, \citenamefont {Mancuso}, \citenamefont {Iboi}, \citenamefont
  {Phan}, \citenamefont {Eikenberry}, \citenamefont {Kuang}, \citenamefont
  {Kostelich},\ and\ \citenamefont {Gumel}}]{eikenberry2020mask}%
  \BibitemOpen
  \bibfield  {author} {\bibinfo {author} {\bibfnamefont {S.~E.}\ \bibnamefont
  {Eikenberry}}, \bibinfo {author} {\bibfnamefont {M.}~\bibnamefont {Mancuso}},
  \bibinfo {author} {\bibfnamefont {E.}~\bibnamefont {Iboi}}, \bibinfo {author}
  {\bibfnamefont {T.}~\bibnamefont {Phan}}, \bibinfo {author} {\bibfnamefont
  {K.}~\bibnamefont {Eikenberry}}, \bibinfo {author} {\bibfnamefont
  {Y.}~\bibnamefont {Kuang}}, \bibinfo {author} {\bibfnamefont
  {E.}~\bibnamefont {Kostelich}}, \ and\ \bibinfo {author} {\bibfnamefont
  {A.~B.}\ \bibnamefont {Gumel}},\ }\bibfield  {title} {\enquote {\bibinfo
  {title} {To mask or not to mask: Modeling the potential for face mask use by
  the general public to curtail the covid-19 pandemic},}\ }\href@noop {}
  {\bibfield  {journal} {\bibinfo  {journal} {Infectious Disease Modelling}\ }
  (\bibinfo {year} {2020})}\BibitemShut {NoStop}%
\bibitem [{\citenamefont {Balachandar}\ \emph {et~al.}(2020)\citenamefont
  {Balachandar}, \citenamefont {Zaleski}, \citenamefont {Soldati},
  \citenamefont {Ahmadi},\ and\ \citenamefont
  {Bourouiba}}]{BALACHANDAR2020103439}%
  \BibitemOpen
  \bibfield  {author} {\bibinfo {author} {\bibfnamefont {S.}~\bibnamefont
  {Balachandar}}, \bibinfo {author} {\bibfnamefont {S.}~\bibnamefont
  {Zaleski}}, \bibinfo {author} {\bibfnamefont {A.}~\bibnamefont {Soldati}},
  \bibinfo {author} {\bibfnamefont {G.}~\bibnamefont {Ahmadi}}, \ and\ \bibinfo
  {author} {\bibfnamefont {L.}~\bibnamefont {Bourouiba}},\ }\bibfield  {title}
  {\enquote {\bibinfo {title} {Host-to-host airborne transmission as a
  multiphase flow problem for science-based social distance guidelines},}\
  }\href {\doibase https://doi.org/10.1016/j.ijmultiphaseflow.2020.103439}
  {\bibfield  {journal} {\bibinfo  {journal} {International Journal of
  Multiphase Flow}\ }\textbf {\bibinfo {volume} {132}},\ \bibinfo {pages}
  {103439} (\bibinfo {year} {2020})}\BibitemShut {NoStop}%
\bibitem [{\citenamefont {Prather}, \citenamefont {Wang},\ and\ \citenamefont
  {Schooley}(2020)}]{prather2020reducing}%
  \BibitemOpen
  \bibfield  {author} {\bibinfo {author} {\bibfnamefont {K.~A.}\ \bibnamefont
  {Prather}}, \bibinfo {author} {\bibfnamefont {C.~C.}\ \bibnamefont {Wang}}, \
  and\ \bibinfo {author} {\bibfnamefont {R.~T.}\ \bibnamefont {Schooley}},\
  }\bibfield  {title} {\enquote {\bibinfo {title} {Reducing transmission of
  sars-cov-2},}\ }\href@noop {} {\bibfield  {journal} {\bibinfo  {journal}
  {Science}\ } (\bibinfo {year} {2020})}\BibitemShut {NoStop}%
\bibitem [{Note5()}]{Note5}%
  \BibitemOpen
  \bibinfo {note} {The relation between $T_d $ and $R_t$ as shown in Fig.~\ref
  {fig_Td} \protect \color {black} means that attempting to describe a wide
  area's current situation by means of some average value of $R_t$ might be
  highly unappropriated. On the other hand, the spread of the disease over
  smaller, more homogeneous areas in which the population shares a certain
  degree of mobility and social behavior would be quite well described by this
  parameter.}\BibitemShut {Stop}%
\bibitem [{Note6()}]{Note6}%
  \BibitemOpen
  \bibinfo {note} {Note that each run starts with the same initial condition,
  only one sick person. Since the Monte Carlo algorithm is based on random
  numbers, we expect that every run will yield slightly different results.
  Repeating the simulation for 100 separated runs evaluates the algorithm's
  robustness. Results have been averaged and analyzed to determine the
  statistical error.}\BibitemShut {Stop}%
\bibitem [{end()}]{endcoronavirus}%
  \BibitemOpen
  \href@noop {} {\bibinfo  {journal}
  {https://www.endcoronavirus.org/countries}\ }\BibitemShut {NoStop}%
\end{thebibliography}%
\end{document}